\documentclass[conference]{IEEEtran}
\IEEEoverridecommandlockouts
\usepackage{cite}
\usepackage{hyperref}
\usepackage{siunitx}
\usepackage{amsmath,amssymb,amsfonts}
\usepackage{algorithmic}
\usepackage{makecell}
\usepackage{multirow}
\ifCLASSINFOpdf
    \usepackage[pdftex]{graphicx}
    \DeclareGraphicsExtensions{.pdf,.jpeg,.png}
\else
  \usepackage[dvips]{graphicx}
\fi

\ifCLASSOPTIONcompsoc
  \usepackage[caption=false,font=normalsize,labelfont=sf,textfont=sf]{subfig}
\else
  \usepackage[caption=false,font=footnotesize]{subfig}
\fi

\usepackage{textcomp}
\usepackage{xcolor}

    \def\Complex{{\rm\rule[.23ex]{.03em}{1.1ex}\kern-.3em{C}}}

    \newcommand{\be}{\begin{equation}} \newcommand{\ee}{\end{equation}}
    \newcommand{\bea}{\begin{eqnarray}} \newcommand{\eea}{\end{eqnarray}}
    \newcommand{\benum}{\begin{enumerate}} \newcommand{\eenum}{\end{enumerate}}

    \usepackage{xcolor}

\def\BibTeX{{\rm B\kern-.05em{\sc i\kern-.025em b}\kern-.08em
    T\kern-.1667em\lower.7ex\hbox{E}\kern-.125emX}}

\begin{document}
\title{Agentic Wireless Digital Twin Construction and Calibration Using Real-World Measurements}

\author{
\IEEEauthorblockN{Fan-Hao~Lin\textsuperscript{*},
 Kun-Li~Shih\textsuperscript{*}, 
 Chu-Hsiang Huang\textsuperscript{†},
 Hui~Chen\textsuperscript{‡}, 
 Chao-Kai~Wen\textsuperscript{*}, 
 Henk Wymeersch\textsuperscript{‡} 
 }

\textsuperscript{*} National Sun Yat-sen University, Taiwan \ \  
\textsuperscript{†} National Taiwan University, Taiwan \\ 
\textsuperscript{‡} Chalmers University of Technology, Sweden 
}
\maketitle

\begin{abstract}
Wireless digital twins (WDTs) are promising enablers for developing and evaluating AI-native radio access networks, yet constructing a high-fidelity WDT typically requires substantial manual effort to integrate heterogeneous information and infer unknown propagation-related parameters. This paper proposes Agentic WDT (AWDT), an end-to-end agentic framework for autonomous WDT construction and calibration using readily available environmental information and measurements readily obtainable from commercial smartphones. AWDT comprises three agents: EnvAgent constructs the propagation environment, OpAgent infers BS and sector configurations, and MatAgent calibrates radio material properties. The agents iteratively refine the WDT using discrepancies between LTE/NR reference signal received power (RSRP) measurements and ray-tracing predictions. Real-world experiments show that AWDT reduces the RSRP prediction MAE from 11.46 to 5.25~dB, demonstrating a substantial improvement in ray-tracing fidelity. Evaluation with an independent measurement system further demonstrates cross-device transferability with lightweight device-specific bias adaptation, highlighting the potential of agentic AI for scalable WDT construction and calibration with limited prior knowledge.
\end{abstract}


\section{Introduction}

With the transition toward sixth-generation mobile communications (6G) and beyond, wireless systems are increasingly evolving toward artificial intelligence (AI)-native radio access networks (AI-RAN), in which AI is expected to support network optimization, control, and autonomous operation \cite{brikAIinORAN}. Developing and validating such AI-driven frameworks requires large-scale and diverse channel data, yet collecting sufficient real-world data through drive tests or dedicated measurement campaigns is costly, time-consuming, and difficult to scale \cite{wang2018channelmeasure}. Wireless digital twins (WDTs) provide a promising alternative by reproducing site-specific channel conditions in a virtual environment, enabling scalable data generation and evaluation for AI-RAN development \cite{khanwdt}. This, however, requires the WDT to accurately reproduce the propagation characteristics of the physical environment.

Ray tracing \cite{aoudia2025sionna} enables site-specific propagation modeling, but its fidelity depends strongly on 3D geometry, operator-side configurations, antenna patterns, and radio materials \cite{wang2025digitaltwinchannel}. Prior work has studied environment reconstruction \cite{chen2024envreconstruct}, BS configuration estimation \cite{eller2022bs}, and radio-material calibration \cite{hoydis2024differentiablert}. Recent agentic frameworks automate radio-map generation and wireless simulation \cite{quan2025llm}, but lack validation against real-world propagation measurements. Measurement-driven WDT calibration has also been explored \cite{ai2026propagation}, but relies on dedicated CSI measurements and a preconstructed environment rather than E2E construction and calibration using measurements readily obtainable from commercial smartphones.

Moreover, the effects of these unknown factors are jointly embedded in the observed channel characteristics, making WDT calibration a highly coupled inverse problem. To address these challenges, we propose Agentic WDT (AWDT), an E2E multi-agent framework for autonomous WDT construction and calibration using commercial smartphone measurements. AWDT coordinates three specialized agents, \emph{Environment Agent (EnvAgent)}, \emph{Operator-side Agent (OpAgent)}, and \emph{Radio Material Agent (MatAgent)}, to progressively construct and calibrate these WDT components through iterative interactions with the ray-tracing engine.
The main contributions of this work are summarized as follows:
\begin{itemize}
    \item We propose AWDT, a goal-driven multi-agent framework for E2E WDT construction and calibration, in which specialized agents operate on physically interpretable WDT components and collaboratively refine propagation-relevant parameters through ray-tracing feedback.

    \item We develop a practical approach for constructing and calibrating WDTs using only publicly available environmental data and measurements readily obtainable from commercial smartphones, without requiring detailed BS configurations or dedicated channel-measurement equipment.

    \item We further evaluate cross-measurement-device transferability using independently collected measurements from another device, demonstrating that the calibrated WDT can be transferred across measurement devices with only lightweight device-specific bias adaptation.
\end{itemize}

\section{Multi-agent Framework For AWDT} \label{Multi-agent Framework}
\begin{figure}[t]
    \centering
    \includegraphics[width=0.95\linewidth]{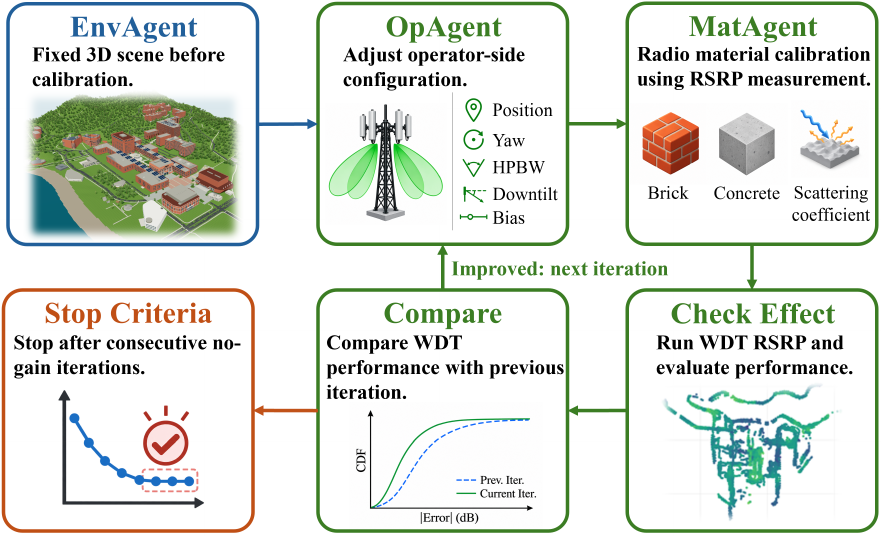}
    \caption{A multi-agent framework for AWDT.}
    \label{fig:framework}
\end{figure}

This section presents a multi-agent framework for AWDT. The framework is shown in Fig.~\ref{fig:framework} and consists of three specialized agents: EnvAgent, OpAgent, and MatAgent. Each agent is responsible for a distinct set of physically interpretable parameters. By leveraging the discrepancies between ray-tracing predictions and real-world measurements, the agents progressively refine a shared WDT in an E2E manner.

\subsection{EnvAgent: Propagation Environment Construction} \label{subsec:EnvAgent}
The EnvAgent constructs the propagation environment by integrating heterogeneous real-world information, including building footprints, building heights, terrain elevation, and other publicly available data. Although such information is widely available, differences in resolution, coverage, representation, and reliability make its alignment and integration into a consistent 3D scene nontrivial. EnvAgent coordinates these complementary sources to construct the propagation environment.

To improve geometric fidelity beyond map-based data, EnvAgent further incorporates visual observations such as street-view and ground-level photographs. These observations provide local structural details that may be missing or inaccurate in public datasets, allowing the agent to refine buildings and other scene geometries.

More importantly, EnvAgent shifts environment construction from a fixed processing pipeline to a goal-driven workflow. Rather than following a predefined set of data sources and processing steps, it determines what information is required for the current scene, selects appropriate sources and tools, and performs further refinement when necessary. The final output is a propagation-ready 3D scene containing the reconstructed terrain and building geometry for subsequent ray-tracing simulation and AWDT calibration. This formulation reduces manual data integration while allowing the construction process to adapt to different levels of data availability without redesigning the entire workflow.

\subsection{OpAgent: Operator-side Configuration Inference} 
\label{subsec:OpAgent}
The OpAgent infers BS locations, antenna heights, and sector parameters, including yaw, horizontal and vertical half-power beamwidths (H-/V-HPBW), downtilt, and sector-specific power bias.\footnote{During OpAgent validation, the sector-specific power bias is estimated as the median sector RSRP residual for candidate comparison. Before MatAgent calibration, it is re-estimated using training samples with low material sensitivity. After material calibration, it is recomputed from the remaining residuals and retained as the final sector-specific power offset.} These parameters are progressively refined using cell identifiers, RSRP measurement distributions, and ray-tracing feedback. This enables third parties without access to detailed operator configurations to construct site-specific WDTs for downstream AI-driven applications, such as connectivity-aware mobility and vehicular edge task offloading.

The BS search is first constrained by grouping cells, identified by their physical cell identities (PCIs), that are likely associated with the same BS.\footnote{For LTE, PCIs sharing the same eNB ID are grouped as co-site cells. Valid NR gNB identifiers and NR cell identity (NCI) are unavailable in our dataset, so the target NR PCIs are assumed to be co-located.} For each sector $p$, OpAgent computes an RSRP-weighted centroid
\begin{equation}
\mathbf{c}_{p}
=
\frac{\sum_{i\in\mathcal{S}_{p}} P_i\mathbf{x}_i}
{\sum_{i\in\mathcal{S}_{p}} P_i},
\qquad
\mathcal{S}_{p}=\{i \mid \mathrm{PCI}_i=p\},
\end{equation}
where $\mathbf{x}_i$ is the measurement location and $P_i$ is the corresponding RSRP in the linear power scale.

For a BS associated with $N_P$ co-located sectors, the corresponding centroids are averaged to obtain a coarse spatial anchor
\begin{equation}
\mathbf{c}_{\mathrm{A}}
=
\frac{1}{N_P}\sum_{p=1}^{N_P}\mathbf{c}_{p}.
\end{equation}
Nearby rooftops around $\mathbf{c}_{\mathrm{A}}$ are then selected as candidate BS sites.

Each candidate BS is initialized with the standardized sector antenna pattern specified in 3GPP TR 38.901 \cite{3gpp_tr38901}, with sector $p$ initially directed toward $\mathbf{c}_p$. OpAgent then refines the BS location and antenna height through ray-tracing evaluation. For measurement point $i$, the RSRP residual is
\begin{equation}
\label{eq:rsrp_res}
E_i=P_i^{\mathrm{dB}}-\widehat{P}_i^{\mathrm{dB}},
\end{equation}
where $P_i^{\mathrm{dB}}$ and $\widehat{P}_i^{\mathrm{dB}}$ are the measured and ray-tracing-predicted RSRP, respectively. Candidate configurations are evaluated based on their agreement with the measurements.

After locating the BS, OpAgent examines the current spatial RSRP residuals and autonomously determines which sector parameter should be refined and in which direction. For example, a yaw mismatch shifts the predicted high-gain region, H-HPBW affects its spatial coverage width, while downtilt and V-HPBW mainly influence residuals across UE distances and elevation angles. These signatures guide \emph{context-driven} refinement, with updates retained only when they improve agreement with the measured RSRP.

\subsection{MatAgent: Radio Material Property Calibration}
\label{subsec:MatAgent}

Calibrating radio materials from commercial RSRP measurements is an underconstrained inverse problem, since each measurement aggregates multiple propagation contributions and different material settings can produce similar received powers. MatAgent addresses this ambiguity through spatially prioritized, simulation-tested updates.

Each building is treated as a calibration unit with relative permittivity $\epsilon_r$, conductivity $\sigma$, and scattering coefficient $S$. To prioritize buildings, MatAgent evaluates the local RSRP mismatch using a distance-weighted building score
\begin{equation}
q_b
=
\left|
\frac{\sum_{i\in\mathcal{D}_{\mathrm{tr}}} w_{i,b}E_i}
{\sum_{i\in\mathcal{D}_{\mathrm{tr}}} w_{i,b}}
\right|,
\qquad
w_{i,b}
=
\exp\left(-\frac{d_{i,b}^2}{2\ell^2}\right),
\end{equation}
where $\mathcal{D}_{\mathrm{tr}}$ is the training set, $E_i$ is the RSRP residual defined in~\eqref{eq:rsrp_res}, $d_{i,b}$ is the shortest distance from measurement point $i$ to the footprint of building $b$, and $\ell=90$ controls the spatial weighting scale. A larger $q_b$ indicates a stronger local mismatch and therefore a higher calibration priority.

Buildings are considered sequentially in descending order of $q_b$, without predefining a fixed number of buildings to calibrate. For each building, MatAgent examines the RSRP residuals and current material settings to determine which of $\epsilon_r$, $\sigma$, or $S$ should be refined and in which direction. Each update is evaluated through ray tracing and retained only if it improves the validation objective. After an accepted update, the residuals and building scores are recomputed before the next calibration action. The process terminates when the validation-based stopping criterion is reached.

\section{Experimental Setup and Results} \label{Experiment Result} 

\subsection{Experiment Setup}
\subsubsection{Experimental Scene}
\begin{figure}[t]
    \centering
    \subfloat[]{
        \begin{minipage}[c]{0.45\linewidth}
            \centering
            \includegraphics[width=0.95\linewidth]{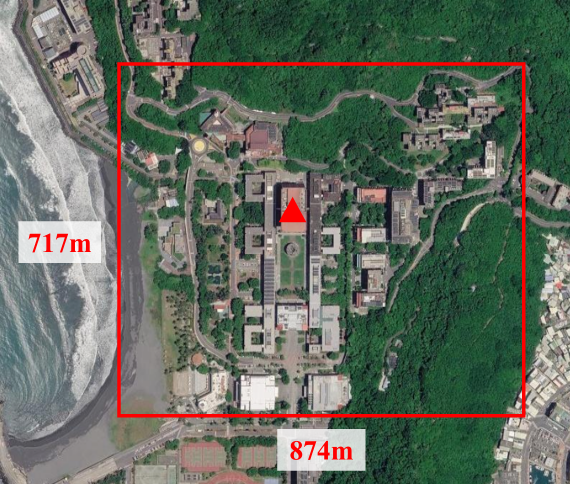}
        \end{minipage}
        \label{fig:NSYSU}
    }
    \hfill
    \subfloat[]{
        \begin{minipage}[c]{0.45\linewidth}
            \centering
            \includegraphics[width=0.95\linewidth]{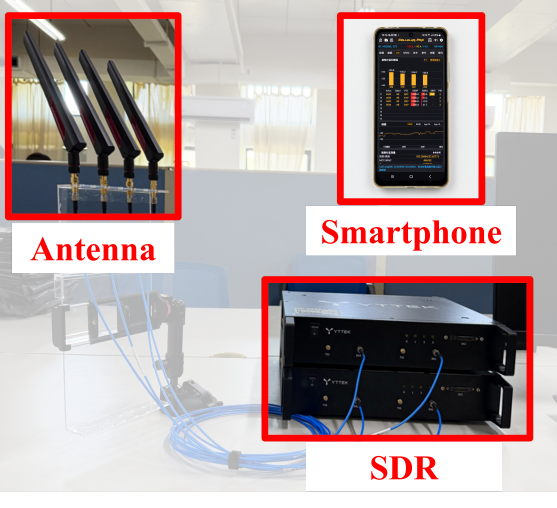}
        \end{minipage}
        \label{fig:measurement_devices}
    }
    \caption{Experimental area and measurement setup.
    (a) NSYSU campus covering approximately
    \SI{717}{m}$\times$\SI{874}{m}, where the red triangle indicates
    the approximate BS site.
    (b) Commercial smartphone and SDR-based measurement setups used for
    cellular RSRP measurements and cross-device evaluation, respectively.}
    \label{fig:measurement_setup}
\end{figure}

The experiments were conducted on the campus of National Sun Yat-sen University (NSYSU), Taiwan. As shown in Fig.~\ref{fig:measurement_setup}(a), the experimental area covers approximately \SI{717}{m}$\times$\SI{874}{m} and includes both flat and mountainous terrain with noticeable elevation variations. The presence of multiple buildings, roads, and open spaces creates diverse propagation conditions, including both line-of-sight (LoS) and non-line-of-sight (NLoS) regions.

\subsubsection{Commercial RSRP measurement} \label{subsubsec:rsrp_measurement}
\begin{figure*}[t]
    \centering

    \subfloat[Real-World LTE B3 RSRP measurement]{
        \includegraphics[width=0.21\linewidth]{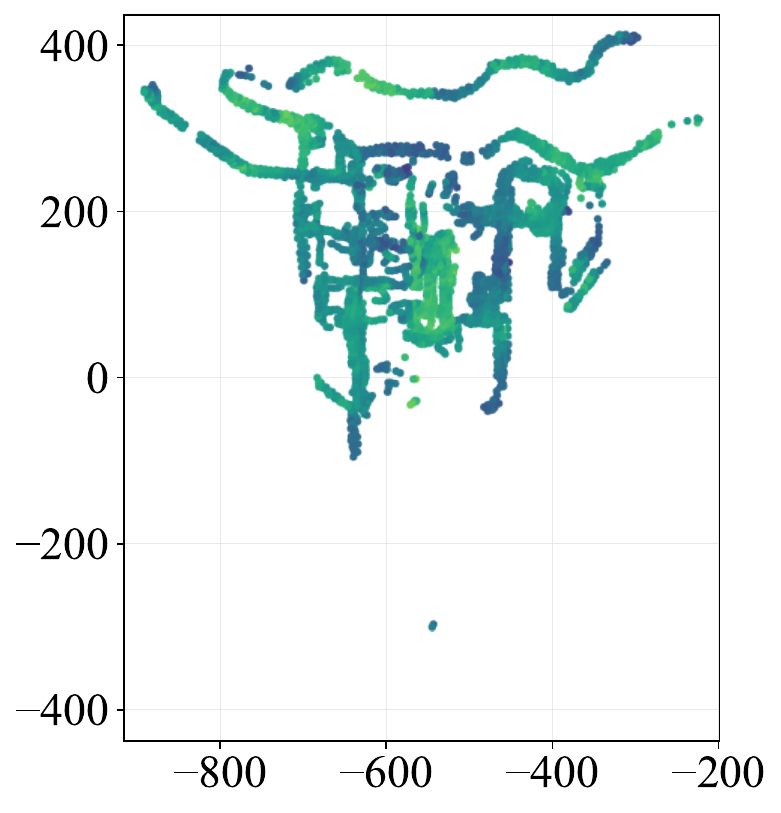}
        \label{fig:real_lte_rsrp}
    }
    \hspace{0.01\linewidth}
    \subfloat[OSM-based scene LTE B3 RSRP map]{
        \includegraphics[width=0.21\linewidth]{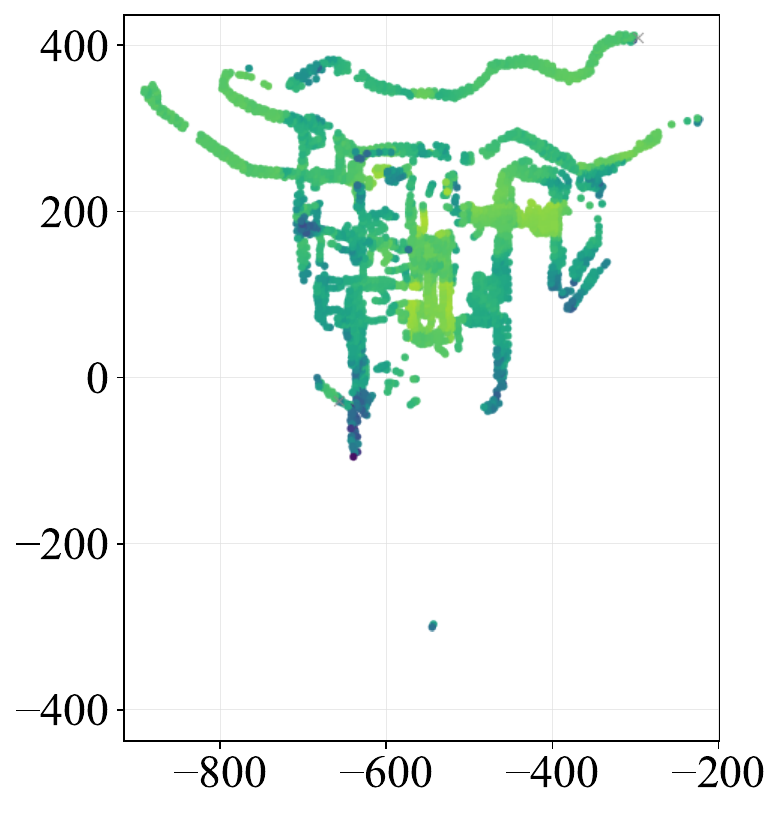}
        \label{fig:osm_lte_rsrp}
    }
    \hspace{0.01\linewidth}
    \subfloat[LTE B3 RSRP map after EnvAgent construction]{
        \includegraphics[width=0.21\linewidth]{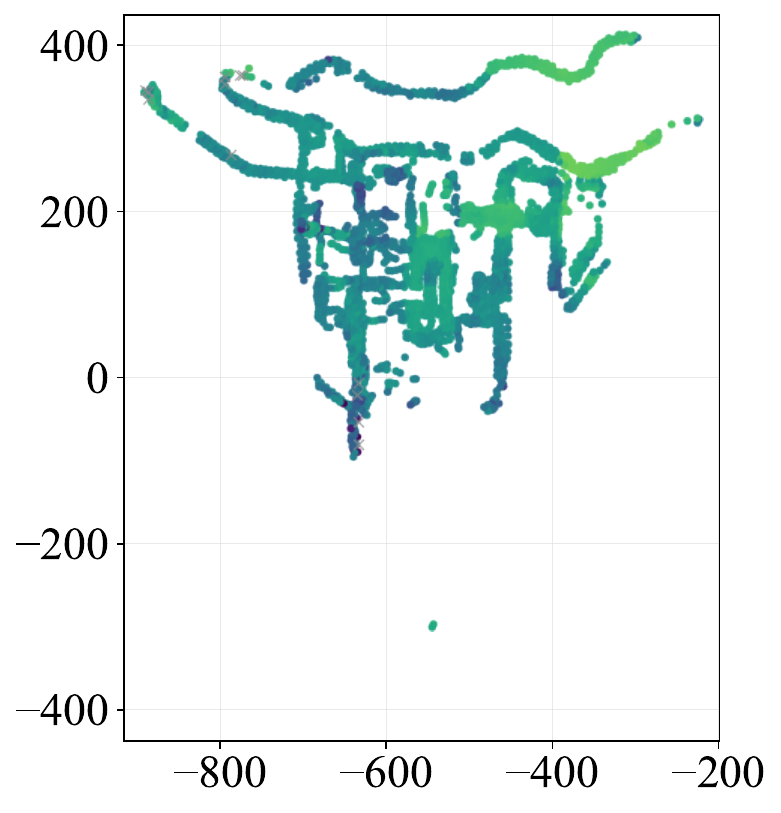}
        \label{fig:updated_lte_rsrp}
    }
    \hspace{0.01\linewidth}
    \subfloat[Calibrated AWDT LTE B3 RSRP map]{
        \includegraphics[width=0.21\linewidth]{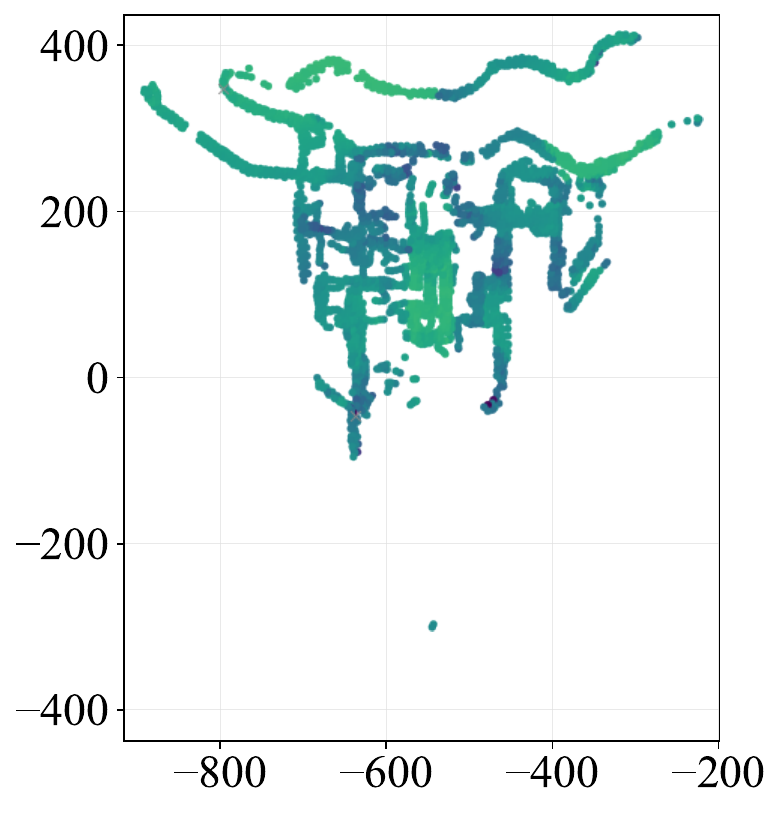}
        \label{fig:calibration_lte_rsrp}
    }

    \vspace{2mm}

    \subfloat[Real-World NR n78 RSRP measurement]{
        \includegraphics[width=0.21\linewidth]{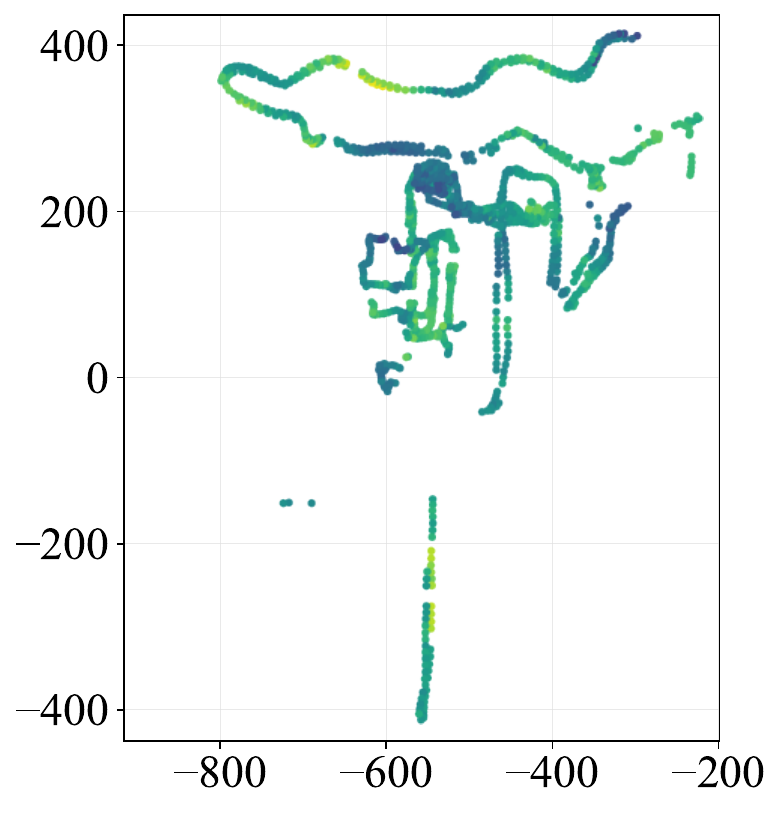}
        \label{fig:real_nr_rsrp}
    }
    \hspace{0.01\linewidth}
    \subfloat[OSM-based scene NR n78 RSRP map]{
        \includegraphics[width=0.21\linewidth]{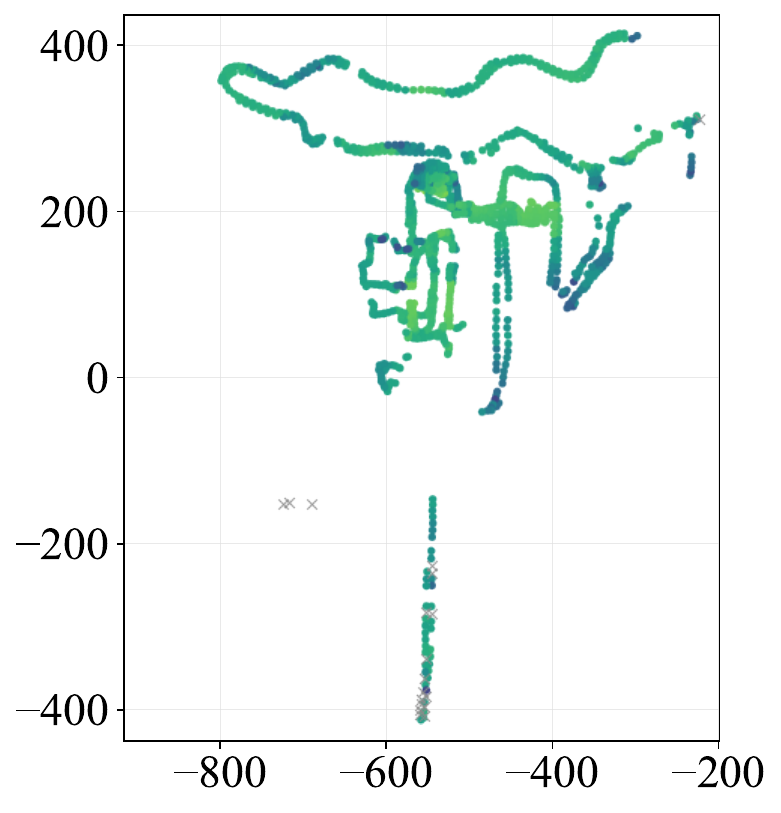}
        \label{fig:osm_nr_rsrp}
    }
    \hspace{0.01\linewidth}
    \subfloat[NR n78 RSRP map after EnvAgent construction]{
        \includegraphics[width=0.21\linewidth]{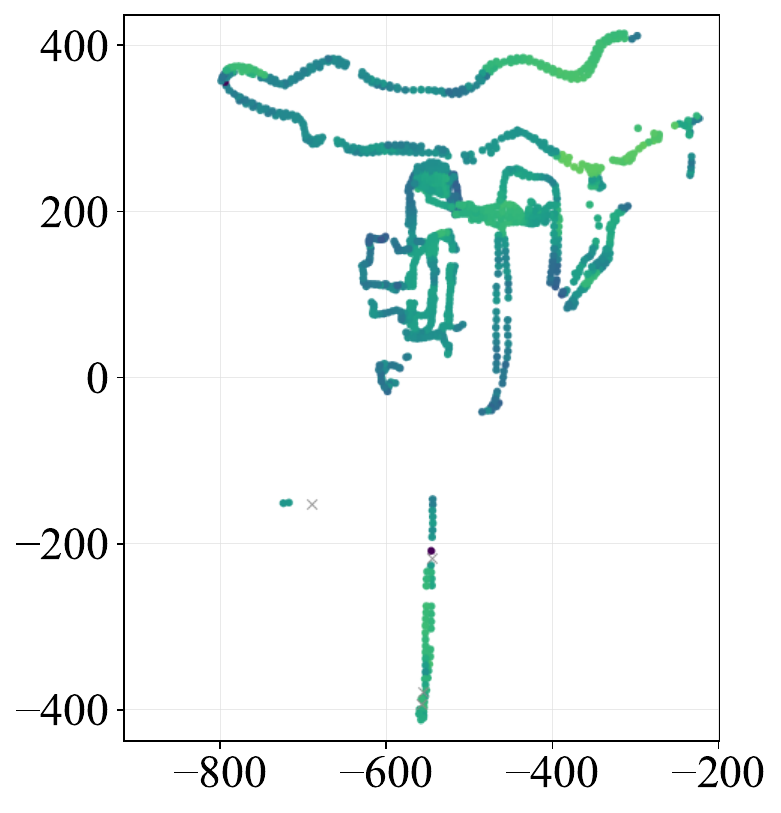}
        \label{fig:updated_nr_rsrp}
    }
    \hspace{0.01\linewidth}
    \subfloat[Calibrated AWDT NR n78 RSRP map]{
        \includegraphics[width=0.21\linewidth]{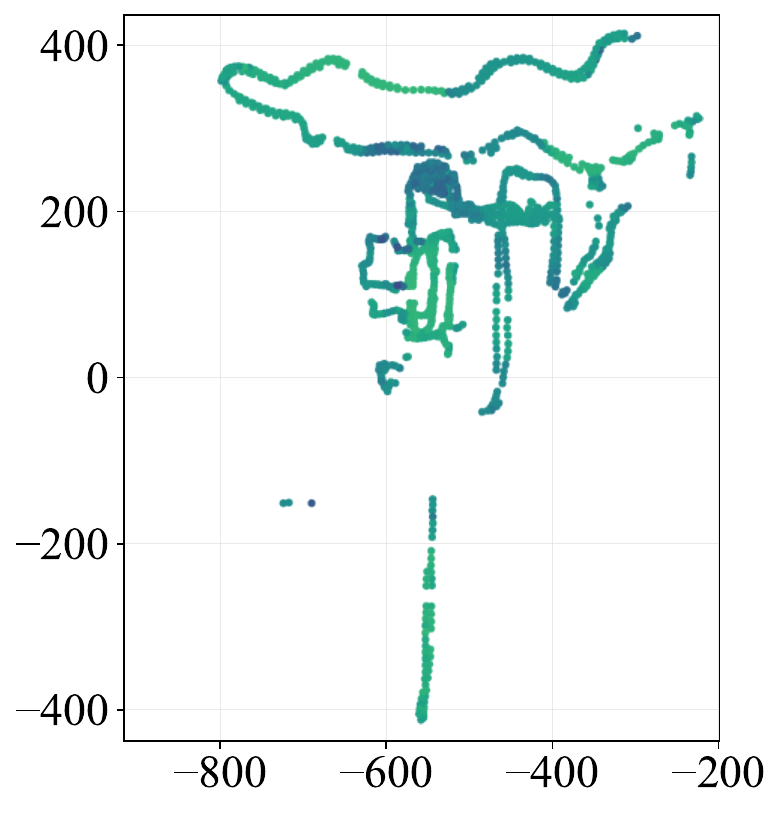}
        \label{fig:calibration_nr_rsrp}
    }

    \includegraphics[width=0.55\textwidth]{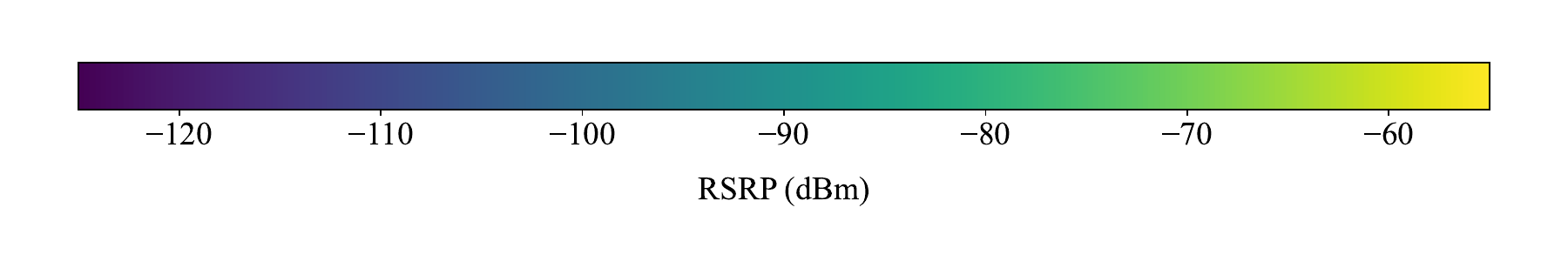}
    \vspace{-0.5cm}

    \caption{Commercial smartphone measurements and ray-tracing-based RSRP maps at different stages of AWDT construction and calibration. The first row presents LTE B3 RSRP, while the second row presents NR n78 SS-RSRP. From left to right, the columns show the real-world measurements, the RSRP map using the initial OSM-based scene, the RSRP map after propagation environment construction, and the RSRP map obtained from the calibrated AWDT using OpAgent and MatAgent. Gray crosses indicate no-path samples for which the ray-tracing engine cannot identify a valid propagation path.}
    \label{fig:rsrp_map}
\end{figure*}

As shown in Fig.~\ref{fig:measurement_setup}(b), we conducted real-world cellular measurements using a commercial smartphone running Cellular-Pro to collect LTE B3 (1.86~GHz) RSRP and NR n78 (3.46~GHz) SS-RSRP, together with their corresponding GPS locations, for AWDT construction and calibration. Fig.~\ref{fig:rsrp_map}(a) and Fig.~\ref{fig:rsrp_map}(e) show the measured LTE RSRP and NR SS-RSRP, respectively.

Since smartphone-reported RSRP may contain device-dependent variations due to differences in RF front-ends and measurement implementations, NR SS-RSRP was additionally collected using a software-defined radio (SDR)-based sniffer with a fixed and explicitly configured receiver gain. The measurement setup is also shown in Fig.~\ref{fig:measurement_setup}(b). The two devices are used to evaluate whether the calibrated AWDT captures transferable propagation characteristics rather than device-specific effects.

\subsubsection{RSRP Data Preprocessing}
Across 34 measurement runs, the RSRP measurements are aggregated into \SI{5}{m}$\times$\SI{5}{m} grid cells, yielding 4,228 occupied cells. To ensure spatial separation, road segments rather than individual cells are assigned to the training, validation, and testing sets, resulting in 3,382 (80\%), 423 (10\%), and 423 (10\%) cells, respectively. The median RSRP in each cell is used to reduce variations caused by smartphone orientation and short-term fluctuations. Training cells are used for calibration, validation cells for update selection, and testing cells for final evaluation.

\subsubsection{Agent Platform}
EnvAgent, OpAgent, and MatAgent are implemented using OpenAI Codex as the underlying agent environment, which supports context-aware reasoning, code execution, and external tool interaction. The role-specific instructions, accessible tools, and AWDT construction and calibration workflows are defined by the proposed framework.

\subsubsection{Ray-Tracing Configuration}
All WDT predictions are generated using NVIDIA Sionna RT \cite{sionna2025}, with reflection, diffraction, scattering, and refraction enabled and a maximum interaction depth of 10. The same configuration is used throughout all construction and calibration stages.

\begin{figure*}[t]
    \centering

    \subfloat[OSM-based scene LTE B3 RSRP error]{
        \includegraphics[width=0.21\linewidth]{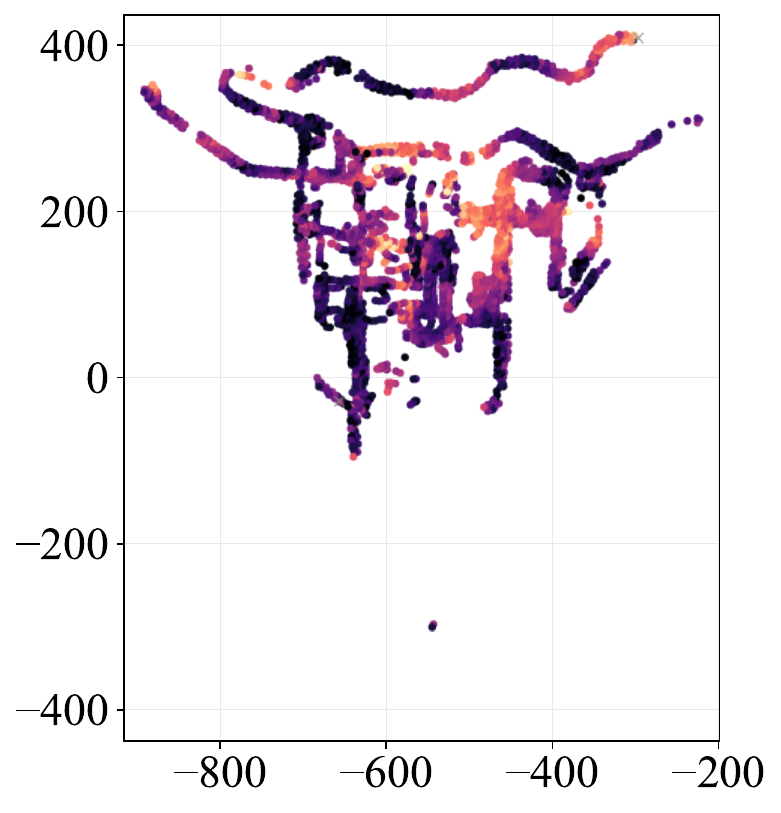}
        \label{fig:osm_lte_rsrp_error}
    }
    \hspace{0.01\linewidth}
    \subfloat[LTE B3 RSRP error after EnvAgent construction]{
        \includegraphics[width=0.21\linewidth]{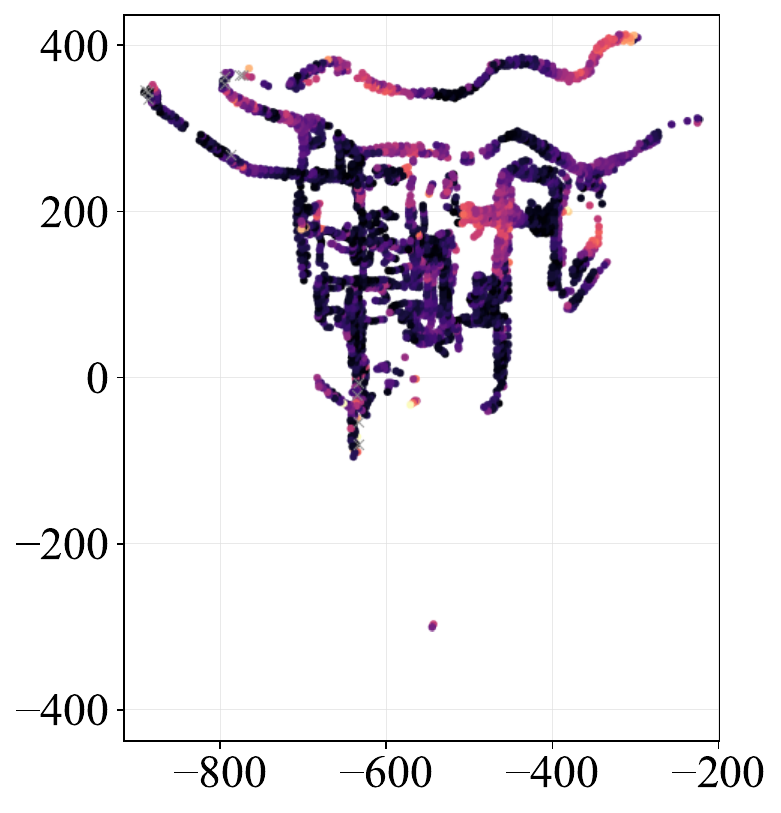}
        \label{fig:updated_lte_rsrp_error}
    }
    \hspace{0.01\linewidth}
    \subfloat[Calibrated AWDT LTE B3 RSRP error]{
        \includegraphics[width=0.21\linewidth]{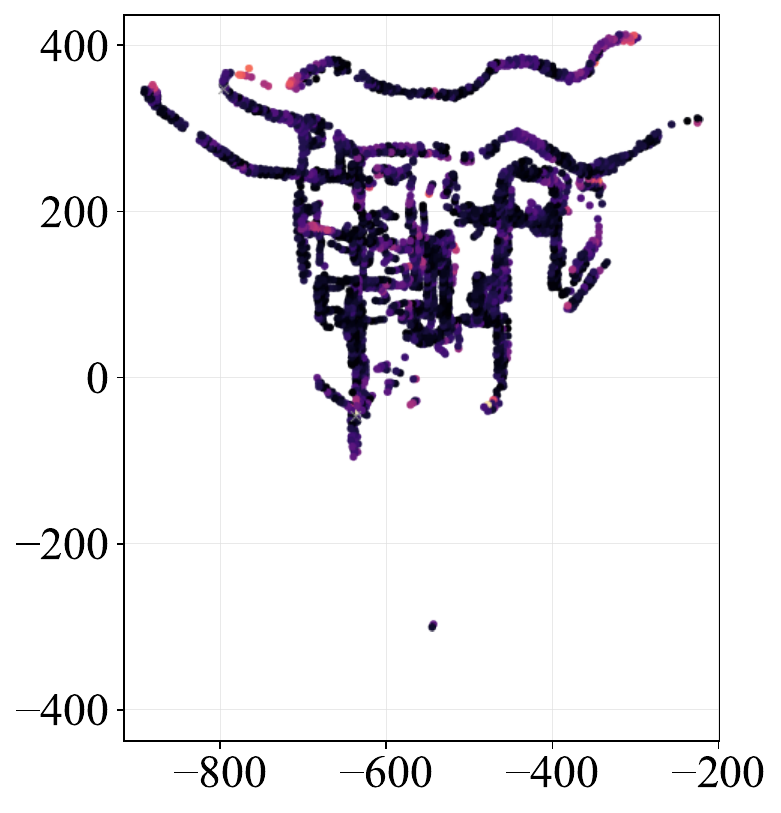}
        \label{fig:calibration_lte_rsrp_error}
    }
    \hspace{0.01\linewidth}
    \subfloat[U-Net LTE B3 RSRP error \cite{li2024geo2sigmap}]{
        \includegraphics[width=0.21\linewidth]{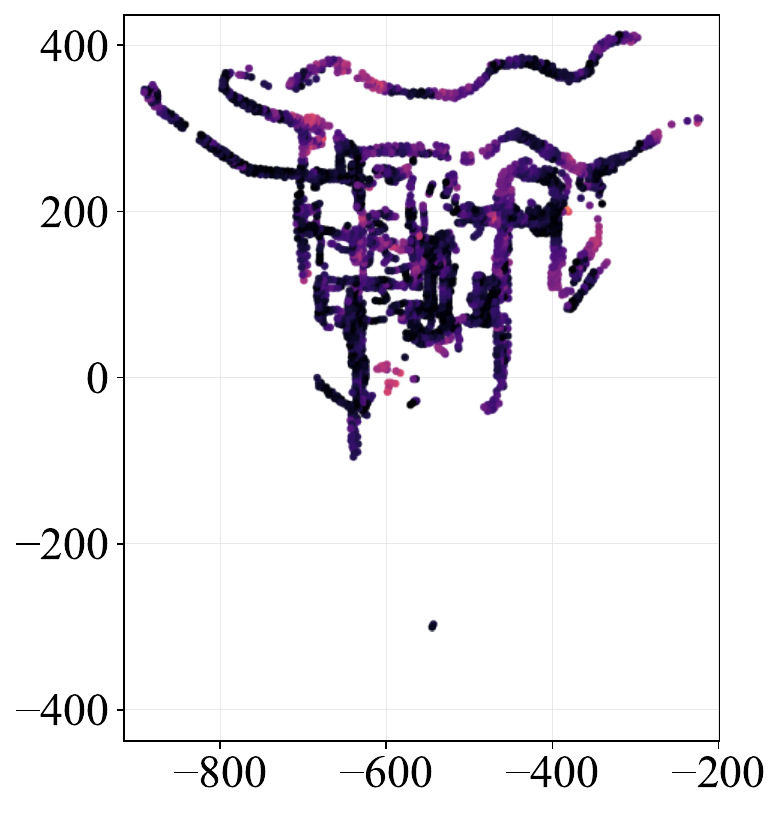}
        \label{fig:unet_lte_rsrp_error}
    }

    \vspace{2mm}

    \subfloat[OSM-based scene NR n78 RSRP error]{
        \includegraphics[width=0.21\linewidth]{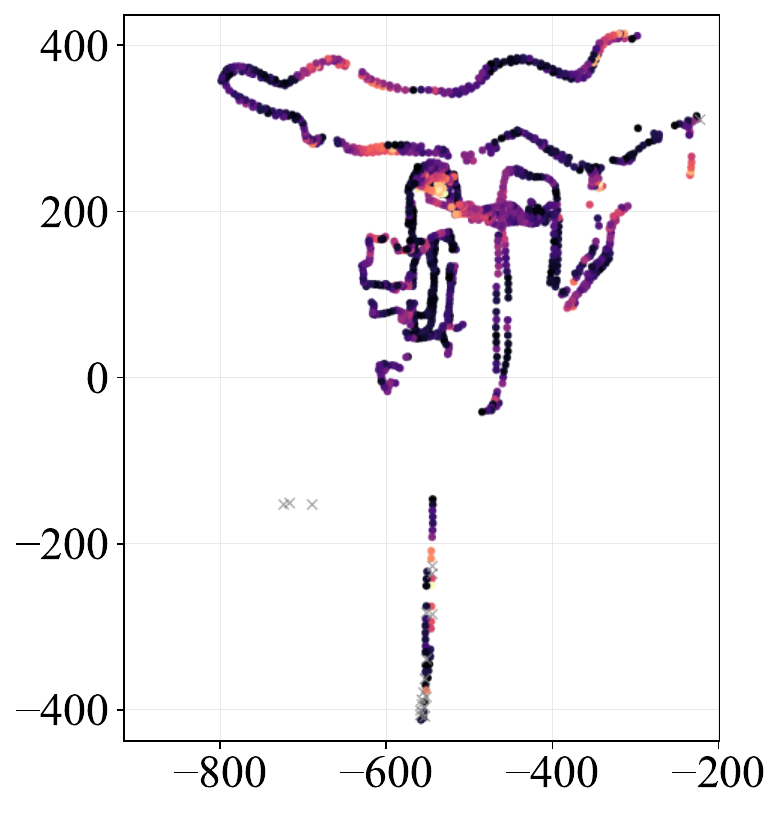}
        \label{fig:osm_nr_rsrp_error}
    }
    \hspace{0.01\linewidth}
    \subfloat[NR n78 RSRP error after EnvAgent construction]{
        \includegraphics[width=0.21\linewidth]{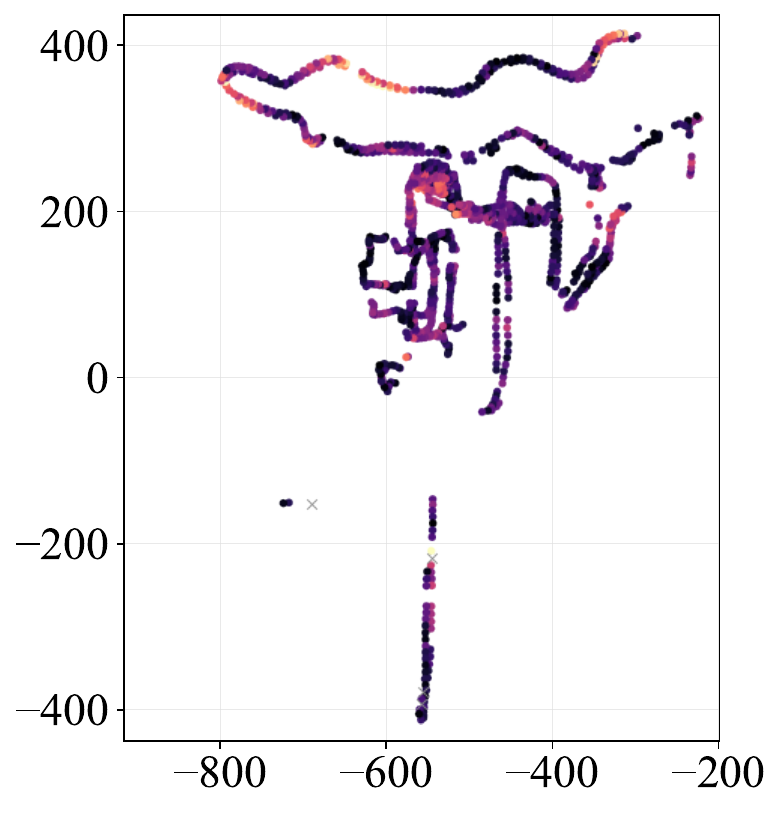}
        \label{fig:updated_nr_rsrp_error}
    }
    \hspace{0.01\linewidth}
    \subfloat[Calibrated AWDT NR n78 RSRP error]{
        \includegraphics[width=0.21\linewidth]{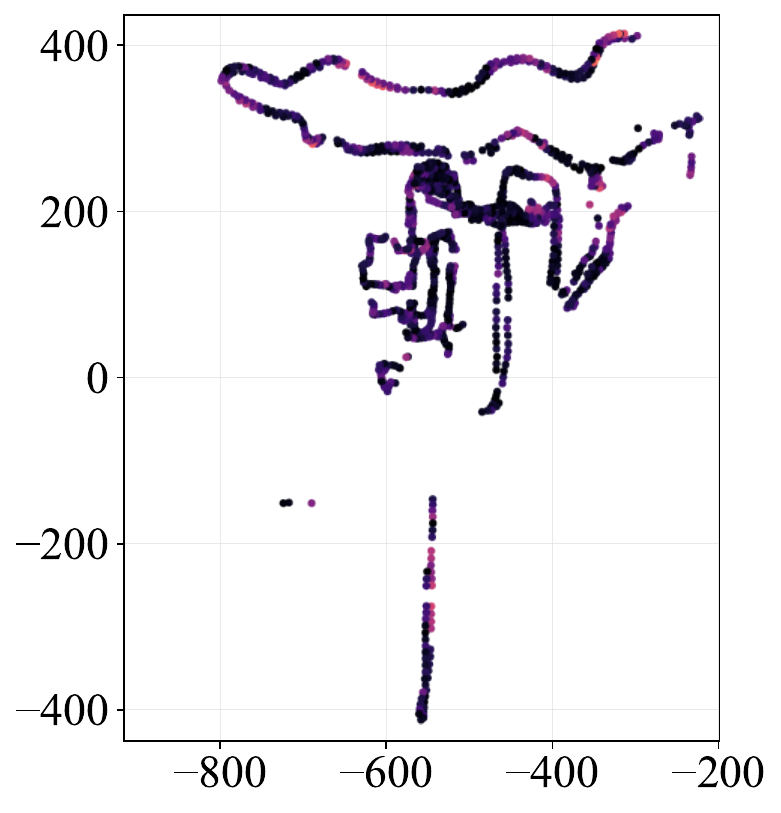}
        \label{fig:calibration_nr_rsrp_error}
    }
    \hspace{0.01\linewidth}
    \subfloat[U-Net NR n78 RSRP error \cite{li2024geo2sigmap}]{
        \includegraphics[width=0.21\linewidth]{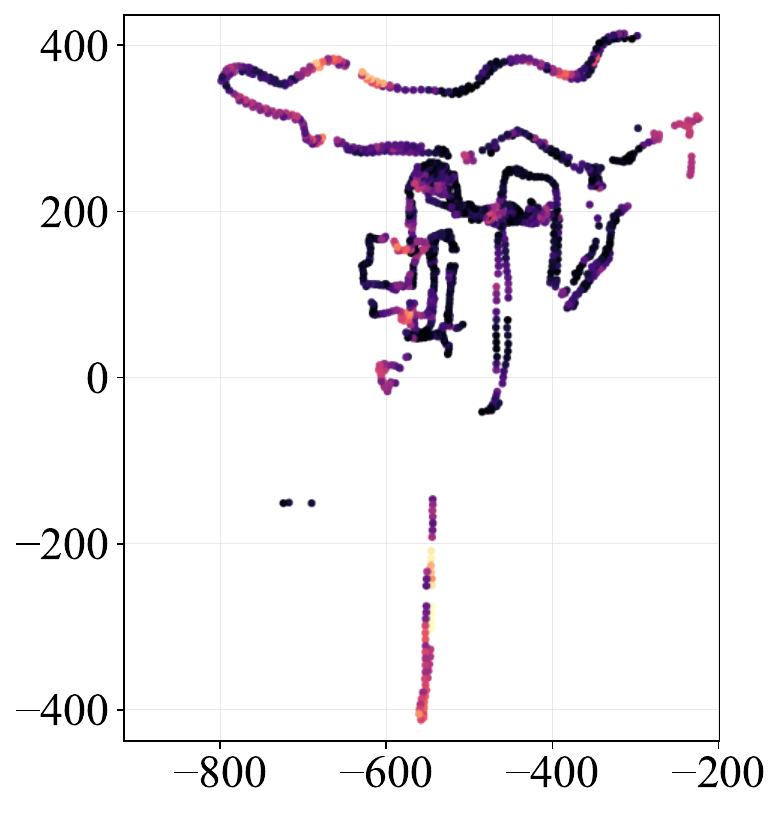}
        \label{fig:unet_nr_rsrp_error}
    }

    \includegraphics[width=0.55\textwidth]{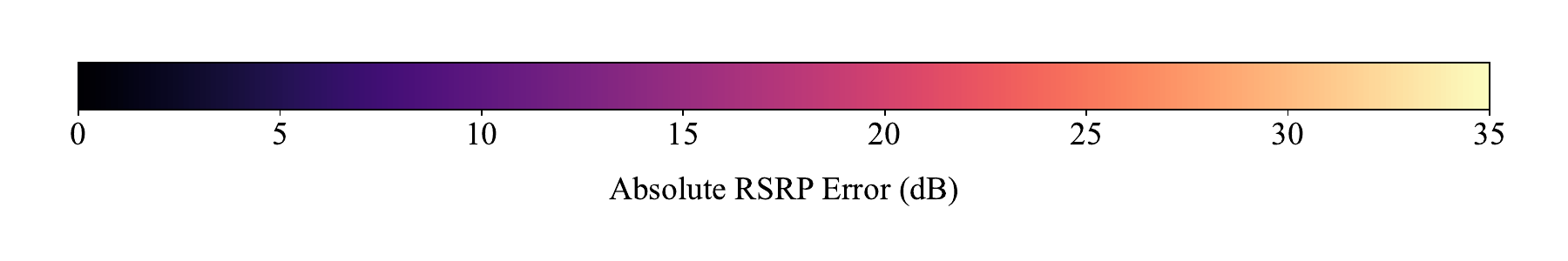}
    \vspace{-0.5cm}

    \caption{Absolute errors between commercial smartphone measurements and ray-tracing predictions at different stages of AWDT construction and calibration. The first row presents LTE B3 RSRP errors, while the second row presents NR n78 SS-RSRP errors. From left to right, the columns correspond to the initial OSM-based scene, the propagation environment constructed by EnvAgent, the calibrated AWDT using OpAgent and MatAgent, and the U-Net-based learning baseline \cite{li2024geo2sigmap}.}
    \label{fig:rsrp_error_map}
\end{figure*}

\subsection{Propagation Environment Construction}
The propagation environment is initialized using building footprints from OpenStreetMap (OSM) \cite{openstreetmap}. However, incomplete building-height information and insufficient terrain representation lead to geometric mismatches with the real environment. EnvAgent incorporates building-height information from the Taiwan 3D Map Service \cite{nlsc_3dmap} and terrain elevation from the NASA SRTM Global 1 Arc Second dataset \cite{nasa_srtm}.

Reference photographs are further used for geometric refinement. By comparing visual observations with the current 3D model, EnvAgent identifies and corrects missing or inaccurate structures, such as corridors and building-specific geometries, to better reproduce the actual propagation environment.

Fig.~\ref{fig:rsrp_map}(b) and Fig.~\ref{fig:rsrp_map}(f) show the ray-tracing-predicted RSRP based on the OSM-based scene. Compared with the measured RSRP, the OSM-based scene substantially overestimates LTE RSRP and NR SS-RSRP, mainly because of underestimated building heights and missing terrain blockage, as highlighted by the error maps in Fig.~\ref{fig:rsrp_error_map}(a) and Fig.~\ref{fig:rsrp_error_map}(e). After EnvAgent refinement, the corresponding results in Fig.~\ref{fig:rsrp_map}(c) and Fig.~\ref{fig:rsrp_map}(g) become more consistent with the measurements, while the error maps in Fig.~\ref{fig:rsrp_error_map}(b) and Fig.~\ref{fig:rsrp_error_map}(f) further demonstrate the improvement in WDT fidelity.

\begin{figure}[t]
    \centering

    \begin{minipage}[c]{0.38\linewidth}
        \centering
        \includegraphics[width=\linewidth]{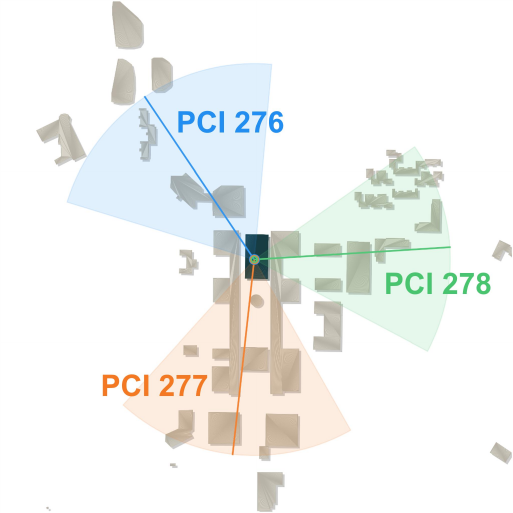}
    \end{minipage}
    \hfill
    \begin{minipage}[c]{0.59\linewidth}
        \centering
        \scriptsize
        \setlength{\tabcolsep}{1.4pt}
        \renewcommand{\arraystretch}{0.95}

        \begin{tabular}{c|c|ccc}
            \hline
            \rule{0pt}{2ex}
            \textbf{Band} &
            \textbf{Parameter} &
            \textbf{PCI 276} &
            \textbf{PCI 277} &
            \textbf{PCI 278} \\
            \hline

            \multirow{5}{*}{LTE B3}
                & Height   & 14.5 m & 3 m & 4 m \\
                & Yaw      & $130.7^\circ$ & $260.8^\circ$ & $8.34^\circ$ \\
                & H-HPBW   & $67^\circ$ & $61^\circ$ & $53^\circ$ \\
                & V-HPBW   & $29^\circ$ & $45^\circ$ & $37^\circ$ \\
                & Downtilt & $10^\circ$ & $7.75^\circ$ & $9.75^\circ$ \\
            \hline

            \multirow{5}{*}{NR n78}
                & Height   & 9 m & 4.5 m & 4 m \\
                & Yaw      & $123.9^\circ$ & $263.7^\circ$ & $3.6^\circ$ \\
                & H-HPBW   & $78^\circ$ & $71.5^\circ$ & $63^\circ$ \\
                & V-HPBW   & $29^\circ$ & $52^\circ$ & $37^\circ$ \\
                & Downtilt & $10^\circ$ & $2^\circ$ & $3.5^\circ$ \\
            \hline
        \end{tabular}
    \end{minipage}

    \caption{Operator-side configurations inferred by OpAgent. LTE and NR share the same BS location.}
    \label{fig:inferred_bs_config}
\end{figure}

\subsection{Commercial RSRP Calibration}
In AWDT, calibration is driven by the discrepancy between ray-tracing predictions and commercial RSRP measurements. Building surfaces are initialized with ITU-based radio material models \cite{ITU2015} to provide physically meaningful priors.
As detailed in Section~\ref{Multi-agent Framework}, OpAgent first infers the BS locations and sector-level configurations, including sector orientations, antenna radiation patterns, and downtilt angles. MatAgent then performs building-wise material calibration by refining the relative permittivity, conductivity, and scattering coefficient based on the spatial RSRP residuals.

The calibration proceeds iteratively, with each update evaluated on the validation set using $\mathcal{L}=E_{P68}+0.2E_{P95}$, where $E_{P68}$ and $E_{P95}$ denote the 68th- and 95th-percentile absolute RSRP errors, respectively. The weight of 0.2 is used to prioritize the typical prediction error represented by $E_{P68}$ while retaining sensitivity to large tail errors through $E_{P95}$. An update is retained only if it reduces $\mathcal{L}$, and calibration terminates after five consecutive candidate updates fail to improve the objective.

Fig.~\ref{fig:inferred_bs_config} illustrates the LTE and NR BS locations and sector configurations inferred by OpAgent. Although detailed operator-side information is unavailable, the approximate BS locations can be identified from the photograph in Fig.~\ref{fig:measurement_setup}(a).\footnote{Multiple BSs from different operators are deployed on this building. The inferred BS location is consistent with the observed deployment at the building level; however, operator-specific ground-truth coordinates are unavailable, so a meter-level localization error cannot be reported.} As shown in Fig.~\ref{fig:inferred_bs_config}, the inferred locations are consistent with the observed deployment. Since the exact sector orientations, antenna patterns, and downtilt angles are also unavailable, their inferred values cannot be directly verified. Instead, their validity is assessed by comparing the resulting WDT predictions with the measured RSRP.

A pronounced RSRP underestimation was observed in the mountainside region served by PCI 276 for both LTE and NR. Although the actual antenna height is unavailable, visual observations indicate a clear LoS between the serving antenna and this region. Assigning PCI 276 an antenna height comparable to those of PCI 277 and PCI 278 results in NLoS conditions in the digital twin. OpAgent therefore infers a greater antenna height to recover the observed LoS geometry, which also substantially reduces the RSRP underestimation. Nevertheless, without operator-side ground truth, the inferred height should be interpreted as a physically consistent configuration rather than an exact reconstruction of the deployed antenna height.

The MatAgent calibration results reveal clear variations in relative permittivity and conductivity across buildings, forming distinct effective radio-material groups with characteristics comparable to those of concrete-like and brick-like materials. In contrast, the scattering coefficients remain within a relatively narrow range. This can be attributed to the outdoor propagation environment, where the measured RSRP is largely governed by dominant propagation paths, such as LoS and strong specular reflections, while the contribution of diffuse scattering is less distinguishable from aggregate received-power measurements. Consequently, most of the material-specific variations across buildings are reflected in the dielectric properties rather than in the scattering coefficient.

Fig.~\ref{fig:rsrp_map}(d) and Fig.~\ref{fig:rsrp_map}(h) show the calibrated AWDT RSRP maps for LTE and NR, while Fig.~\ref{fig:rsrp_error_map}(c) and Fig.~\ref{fig:rsrp_error_map}(g) show the corresponding absolute errors. Although EnvAgent improves the spatial distribution of the predicted RSRP, residual discrepancies remain due to unknown operator-side configurations and radio material properties. By further calibrating the BS/sector configurations and radio material properties, OpAgent and MatAgent improve the agreement between ray-tracing predictions and commercial measurements.

\begin{figure}[t]
    \centering

    \includegraphics[width=0.75\linewidth]{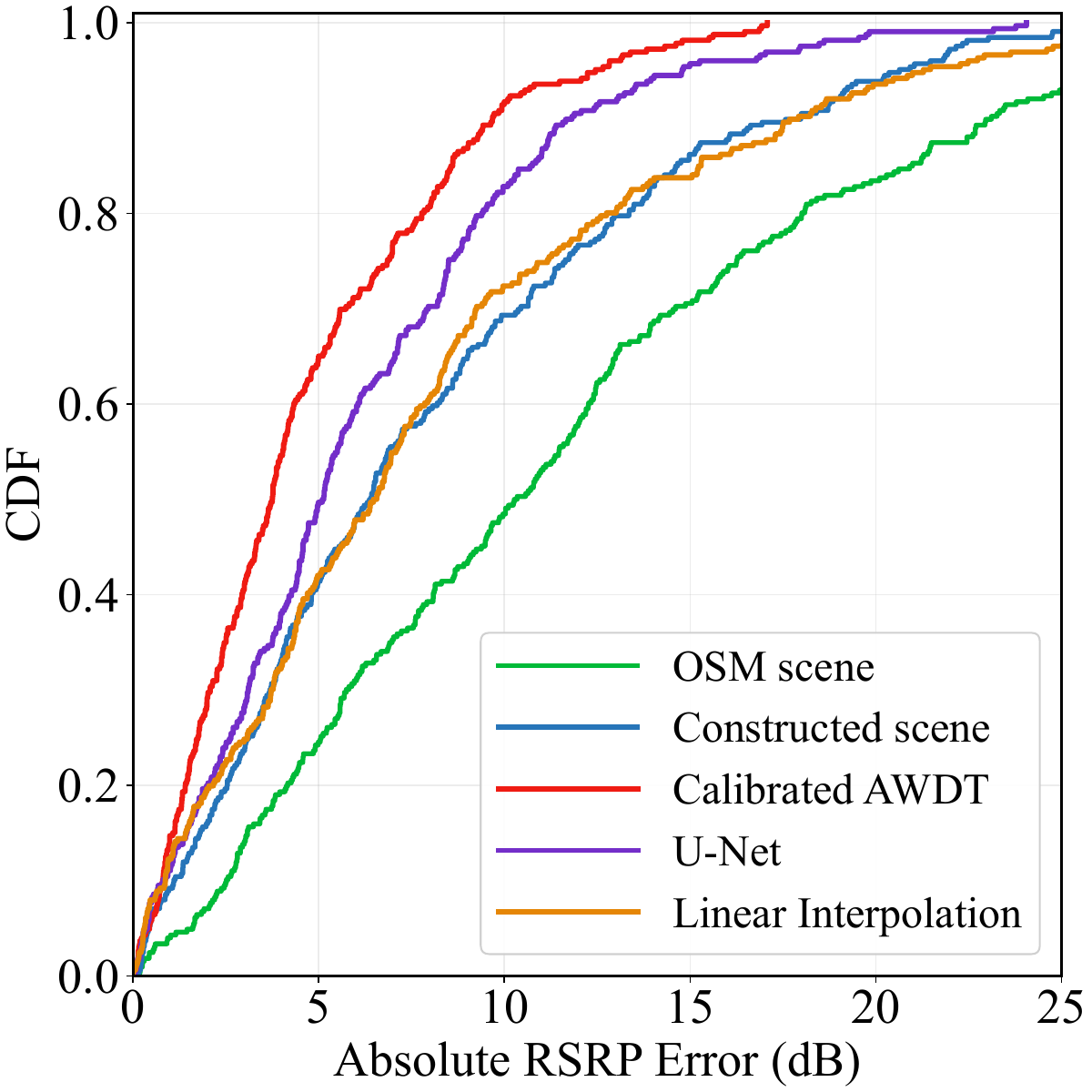}

    \vspace{2mm}

    \footnotesize
    \setlength{\tabcolsep}{2pt}
    \begin{tabular}{lccccc}
    \hline
    \textbf{Metric} &
    \makecell{\textbf{OSM-based} \\ \textbf{Scene}} &
    \makecell{\textbf{Constructed} \\ \textbf{Scene}} &
    \makecell{\textbf{Calibrated} \\ \textbf{AWDT}} &
    \textbf{U-Net}~\cite{li2024geo2sigmap} &
    \makecell{\textbf{Linear} \\ \textbf{Interpolation}} \\
    \hline

    $E_{\rm P68}$ &
    14.29 dB &
    9.70 dB &
    \textbf{6.49 dB} &
    7.42 dB &
    8.99 dB \\

    $E_{\rm P95}$ &
    25.11 dB &
    20.63 dB &
    \textbf{13.77 dB} &
    14.78 dB &
    21.21 dB \\

    MAE &
    11.46 dB &
    8.07 dB &
    \textbf{5.25 dB} &
    6.03 dB &
    7.90 dB \\
    \hline
    \end{tabular}

    \caption{Comparison of RSRP prediction errors at different stages of AWDT construction and calibration for LTE B3 and NR n78 measurements. The CDF illustrates the absolute RSRP error distributions, while the table summarizes the 68th-percentile error $E_{\rm P68}$, 95th-percentile error $E_{\rm P95}$, and mean absolute error (MAE). The same sector-specific median bias is applied to all stages, with no-path samples assigned the absolute measured RSRP as their error.}
    \label{fig:error_cdf}
\end{figure}

To further evaluate AWDT, we compare it with a U-Net-based data-driven baseline \cite{li2024geo2sigmap} and a linear-interpolation baseline. As shown in Fig.~\ref{fig:rsrp_map} and Fig.~\ref{fig:rsrp_error_map}, AWDT achieves lower errors over most of the evaluated area. Fig.~\ref{fig:error_cdf} further summarizes the improvements across different construction and calibration stages, with the MAE decreasing from 11.46~dB for the OSM-based scene to 5.25~dB after calibration, outperforming both baselines. Unlike the learning-based and interpolation baselines, AWDT uses commercial measurements as feedback to refine physically interpretable WDT components, producing a calibrated physical model that can be reused for subsequent ray-tracing-based analyses.

\subsection{Cross-Measurement-Device Evaluation}
RSRP measurements may exhibit device-dependent variations due to differences in RF front-ends and receiver implementations. Since the internal gain states and compensation procedures of commercial smartphones are generally inaccessible, we additionally collect NR SS-RSRP using an SDR-based sniffer with a fixed and explicitly configured receiver gain. The SDR estimates SS-RSRP from the resource elements carrying the secondary synchronization signal, providing an independent reference for evaluating whether the improvements obtained from smartphone-based calibration are preserved when evaluated using an independent measurement device.

The AWDT is first constructed and calibrated using commercial smartphone measurements. During adaptation to the SDR-based sniffer, all propagation-related parameters are kept fixed, while only a device-specific global RSRP bias is fine-tuned using 20\% of the SDR measurements to compensate for hardware-dependent offsets. The remaining SDR measurements are used for evaluation.

Table~\ref{tab:sdr_error} compares the RSRP errors on the SDR measurements at different stages. With only a global device-specific bias correction, the calibrated AWDT achieves consistently lower prediction errors, indicating that the improvements from smartphone-based calibration are largely preserved across measurement devices. The results therefore suggest that the proposed AWDT captures propagation characteristics that are transferable across measurement hardware rather than overfitting to a particular smartphone.

\begin{table}[t]
    \caption{Cross-Measurement-Device Evaluation Using SDR-Based NR SS-RSRP Measurements}
    \label{tab:sdr_error}
    \centering
    \footnotesize
    \setlength{\tabcolsep}{3pt}
    \renewcommand{\arraystretch}{1.1}

    \begin{tabular}{lccc}
        \hline
        \textbf{Metric} &
        \makecell{\textbf{OSM-based} \\ \textbf{Scene}} &
        \makecell{\textbf{Constructed} \\ \textbf{Scene}} &
        \makecell{\textbf{Calibrated} \\ \textbf{AWDT}} \\
        \hline

        $E_{\rm P68}$ &
        11.72 dB &
        10.67 dB &
        \textbf{7.71 dB} \\

        $E_{\rm P95}$ &
        21.64 dB &
        20.90 dB &
        \textbf{13.97 dB} \\

        MAE &
        9.34 dB &
        8.82 dB &
        \textbf{6.07 dB} \\
        \hline
    \end{tabular}
\end{table}

\section{Conclusion} \label{Conclusion}
This work proposed AWDT, an E2E agentic framework for autonomous WDT construction and calibration. EnvAgent, OpAgent, and MatAgent jointly construct the propagation environment, infer operator-side configurations, and calibrate radio material properties using commercial RSRP measurements with limited prior information. AWDT reduces the MAE from 11.46~dB to 5.25~dB and demonstrates cross-device transferability with lightweight device-specific bias adaptation. These results highlight the potential of agentic AI for scalable and practical WDT construction and calibration.

\bibliographystyle{IEEEtran}
\bibliography{references}

\end{document}